\documentclass[preprint,12pt]{aastex}
\def\rhobar{\bar\rho}
\def\delbar{\bar\delta}
\def\k0{\kappa_0}

\begin{document} 

\title{Schechter vs. Schechter: Sub-Arcsecond Gravitational Lensing
and Inner Halo Profiles}

\author{Chung-Pei Ma}
\affil{Department of Astronomy, University of California at Berkeley,
601 Campbell Hall, Berkeley, CA~94720; cpma@astro.berkeley.edu }
              
\begin{abstract}

Sub-arcsecond lensing statistics depend sensitively on the inner mass
profiles of low-mass objects and the faint-end slopes of the Schechter
luminosity function and the Press-Schechter mass function.  By
requiring the luminosity and mass functions to give consistent
predictions for the distribution of image separation below 1'', we
show that dark matter halos with masses below $10^{12} M_\odot$ cannot
have a single type of profile, be it the singular isothermal sphere
(SIS) or the shallower ``universal'' dark matter profile.  Instead,
consistent results are achieved if we allow a fraction of the halos at
a given mass to be luminous with the SIS profile, and the rest to be
dark with an inner logarithmic slope shallower than $-1.5$ to
compensate for the steeper faint-end slope of the mass function
compared with the luminosity function.  We quantify how rapidly the
SIS fraction must decrease with decreasing halo mass, thereby
providing a statistical measure for the effectiveness of feedback
processes on the baryon content in low-mass halos.

\end{abstract}

\keywords{galaxies: evolution -- galaxies: structure -- 
gravitational lensing}

\section{Introduction}

The distribution of the luminosity of galaxies and the distribution of
the mass of dark matter halos are well approximated by the Schechter
luminosity function \citep{s76} and the Press-Schechter mass function
\citep{ps74}, respectively.  Both functions increase as a power law
towards the low luminosity and mass ends, but the mass function
increases with a steeper slope than the luminosity function.  Low mass
halos must therefore contain relatively less luminous baryonic
material in comparison with massive halos.  Detailed models of galaxy
formation have been able to account for this difference by feedback
processes such as supernova explosions, stellar winds, and
photo-ionizations that suppress the amount of baryons and star
formation rates in low mass halos (e.g., Benson et al. 2002;
Somerville \& Primack 1999; Kaufmann et al 1993).

In this Letter we examine this issue from a different perspective of
small-separation strong gravitational lensing.  The image separation
distribution of lenses below 1'' depends sensitively on both the inner
mass profile of galactic halos and the faint end slope of the mass and
luminosity functions.  We compare the traditional approach that models
the lenses as SIS and the Schechter luminosity function, with a
dark-matter based approach that models the lenses with a certain halo
mass profile and the Press-Schechter mass function.  We investigate
the constraints on the inner {\it total} mass profiles of halos by
requiring the two approaches to give consistent predictions.  Since
evidence based on stellar dynamics of elliptical galaxies (e.g., Rix
et al. 1997; Romanowsky \& Kochanek 1999; Treu \& Koopmans 2002),
modeling of lensed systems (e.g., Cohn et al. 2001), and flux ratios
of multiple images \citep{rm01, rusin02} all give an inner profile for
lensing galaxies that is consistent with SIS, we will use SIS in the
lensing calculation with the luminosity function.  Dark matter halos
then clearly cannot all be SIS because if so, the mass function is
steeper than the luminosity function and would lead to a relatively
higher lensing rate at smaller angular scale.  We will show that
modifying the SIS to any {\it single} flatter profile for halos does
not work either.  Instead, we discuss in \S~4 how a combination of
profiles is needed to resolve the problem.

This paper complements several recent studies on strong lensing
statistics in which the emphasis is on the effects of lens mass
profiles and baryon compression on the cumulative lensing rates at
$\theta \ga 1''$ and the implications for cosmological parameters from
the scarcity of large separation ($\ga 3''$) systems \citep{km01,
keeton01,kw01,k01,sarbu01, lo02, oguri02}.  It is pointed out that
modeling cluster-scale lenses with mass profiles shallower than the
SIS greatly reduces the lensing rate and brings the concordance CDM
model predictions into agreement with observations.  The focus here is
on the less explored sub-arcsec range.  We use the predicted shape for
the differential distribution of image separation to quantify how
rapidly the fraction of SIS halos must decrease with decreasing mass.

In this paper the cosmological model is taken to have a present-day
matter density $\Omega_m=0.3$ (with 0.05 in baryons), cosmological
constant $\Omega_\Lambda=0.7$, Hubble parameter $h=0.75$, and matter
fluctuation $\sigma_8=0.92$.  The lens potentials are assumed to be
spherically symmetric because we are mainly concerned with the lensing
optical depth, which is more sensitive to the velocity dispersion and
the radial profile of the lens than its ellipticity \citep{kb87}.  The
luminosity function is assumed to have a constant comoving galaxy
number density, which is consistent with the nearly constant comoving
halo number density (for a fixed velocity) up to redshift $\sim 5$ in
the Press-Schechter formula \citep{bullock01}.

\section{Lensing Rates from Luminosity and Velocity Functions}

The galaxy luminosity function takes the form \citep{s76}
\begin{equation}
	\phi(L) dL = \phi_* \, \left( {L\over L_*} \right) ^\alpha 
		e^{-L/L_*}  d{\left( L\over L_* \right) }\,.
\end{equation}
An alternative measure is given by the (circular) velocity function 
\begin{equation}
	\psi(v) dv = \psi_* \, \left( {v\over v_*} \right) ^\beta
		e^{-(v/v_*)^n}  d{\left( v\over v_* \right) }\,,
\label{vel}
\end{equation}
which is related to the luminosity function by $L\propto v^n$,
$\beta+1=n(\alpha+1)$, and $\psi_*=n\phi_*$.  The velocity function
can be derived from galaxy survey luminosity functions and kinematic
luminosity-velocity relations, e.g., from various large pre-SDSS
optical surveys \citep{G00} and the 2MASS infrared survey \citep{k01}.
The image separation distribution of lenses at angle $\theta$ is
related to the galaxy velocity function by
\begin{equation}
    {dP \over d\theta} =
    \int dz_l \frac{dr}{dz_l} {dv_c \over d\theta}\, \psi(v_c,z_l)
       \, \sigma_{\rm lens}(v_c,z_l)\,  B \,,
\label{Plum}
\end{equation}    
where $dr/dz=cH_0^{-1} (1+z)^{-1} [\Omega_m(1+z)^3 +
\Omega_\Lambda]^{-1/2}$ (for $\Omega_m+\Omega_\Lambda=1$),
$\sigma_{\rm lens}$ is the lensing cross section, $B$ is the
magnification bias, and $\psi(v_c,z_l) dv_c$ is the physical number
density of galaxies with circular velocity between $v_c$ and
$v_c+dv_c$ at lens redshift $z_l$.

An SIS lens has a density profile $\rho(r)=v_c^2/(8\pi G r^2)$ and
produces an image separation of $\theta=2\theta_E$ where
$\theta_E=2\pi(v_c/c)^2 D_{ls}/D_s$ is the Einstein radius.  For
$\theta=1''$ and source redshift $z_s=1.2$, $v_c$ ranges from 225 to
325 km/s for $z_l=0.3$ to 0.7.  The lensing cross section is
$\sigma_{\rm lens}=\pi(\theta_E D_l)^2 = 4\pi^3 (v_c/c)^4
(D_lD_{ls}/D_s)^2$ \citep{SEF92}.  For small $\theta$ ($< 1''$), one
can show analytically that the slope of $dP/d\theta$ in eq.~(3)
depends only on the faint-end slope $\beta$ of the velocity function:
\begin{equation}
  {dP\over d\theta } \propto \theta^{(\beta+3)/2} 
  \propto \theta^{n(\alpha+1)/2+1} \,, \quad {\rm for\ small\ } \theta\,,
\end{equation}
where $\beta+1=n(\alpha+1)$ is used to relate $\beta$ to the faint-end
slope $\alpha$ of the luminosity function and the luminosity-velocity
relation $L\propto v^n$.  The distribution $dP/d\theta$ therefore has
a positive slope on sub-arcsec scale if the velocity function is
shallower than $\beta=-3$, or if the luminosity function is shallower
than $\alpha=-(1+2/n)$.  Fig.~1 illustrates this dependence.  Recent
surveys favor $\beta\sim -1.3$ to $-1.0$ (e.g., Gonzalez et al. 2000;
Kochanek 2001), indicating a positive slope for $dP/d\theta$ at small
$\theta$.  The image separation distribution of 13 lenses found in the
8958 CLASS radio sources \citep{b02} is shown in Fig.~1 for
comparison.  Note that the smallest and largest $\theta$ bins contain only
one lens each.

\section{Lensing Rates from Mass Functions}

Lensing probes mass after all, so let us take an alternative approach
by modeling the lenses as a population of dark matter halos with a
Press-Schechter type of mass function.  Similar to eq.~(\ref{Plum}),
the image separation distribution is given by
\begin{equation}
  {dP \over d\theta} = 
    \int dz_l \frac{dr}{dz_l} {dM\over d\theta} \,
    n(M,z_l)\, \sigma_{\rm lens}(M,z_l)\, B \,,
\label{Pmass}
\end{equation}    
where $n(M,z_l) dM$ is the physical number density of dark halos with
mass between $M$ and $M+dM$ at $z_l$.  We use the improved version of
the mass function by \cite{jenkins}.  Unlike the lensing galaxies in
eq.~(\ref{Plum}), the lensing halos in eq.~(\ref{Pmass}) may or may
not host central baryons depending on if the lens is baryon or dark
matter dominated near its center.  We will therefore consider
different inner mass profiles.  For luminous lenses in which baryon
dissipation controls the inner density, we use the SIS profile as
in \S~2.  For dark lenses without a significant amount of baryons, we
consider the shallower profiles found in high resolution dark
matter simulations: $\rho \propto r^{-1}$ (Navarro et al. 1997) and
$r^{-1.5}$ \citep{Moore99}.

The lensing properties of the three profiles -- SIS, NFW, and Moore --
are as follows.  For SIS, we relate the circular velocity $v_c$ of
galaxies to the virial velocity $v_{vir}$ of dark halos by
$v_c=\gamma_v v_{vir}$, where $\gamma_v \sim 1.3$ to 1.8 from various
baryon compression models and observational constraints (Oguri 2002
and references therein).  We take Oguri's best fit value
$\gamma_v=1.67$ here.  Using $v_{vir}=(4\pi G^3 M^2
\bar\rho\Delta_{vir}/3)^{1/6}$, we then have $\theta=A M^{2/3}$ where
$A=4\pi G (\gamma_v/c)^2
(D_{ls}/D_s)(4\pi\bar\rho\Delta_{vir}/3)^{1/3}$ and
$\Delta_{vir}\approx 178$.

For the Moore et al. profile $\rho(x)= \rhobar\,\delbar / (x^{3/2} +
x^3)$, we find the projected surface density $\Sigma$ well
approximated by $ \kappa(x)= \Sigma/\Sigma_c =5.4\k0 / [x^{1/2}
(1+1.8\,x)^{3/2}]$ where $x=r/r_s$.
Here $\kappa$ is the convergence, $\Sigma_c=(c^2/4\pi G)(D_s/D_l
D_{ls})$ is the critical surface density, and $\k0=r_s
\rhobar\,\delbar/\Sigma_c$.  The scale radius $r_s$ is related to the
concentration parameter by $c \equiv r_{vir}/r_s$, where $r_{vir}$ is
the halo virial radius, and $\bar\delta = 100\,c^3/ \ln(1+c^{3/2})$.
The reduced deflection angle, related to $\kappa$ by $\alpha(x)=2
x^{-1}\int_0^x dy\, y \kappa(y)$, then has the simple analytic form
$\alpha(x) = (12 \k0/ \sqrt{b}\, x) [\ln (\sqrt{bx}+\sqrt{1+bx}) -
{\sqrt{bx}/ \sqrt{1+bx}}]$, where $b=1.8$.  The lensing cross section
is $\sigma_{\rm lens}=\pi (\beta_{rad} D_l)^2$, where $\beta_{rad}$ is
the angular size of the radial caustic, which we obtain by solving the
lens equation.  

For the Moore profile, we find the fitting function
$(\beta_{rad} D_l) = 9.3 r_s \k0^2/ (1-1.1 \k0^{0.4} + 4.5 \k0^{0.9})$
accurate (with $<5$\% error for $\k0 \la 6$) and useful in speeding up
the computation.
Since the angular separation of the outermost images is insensitive to
the location of the source (Schneider et al. 1992), we use the size of
the tangential critical curve for the image separation:
$\theta=2\theta_{tan}$.  We find the fitting function $(\theta_{tan}
D_l) = 38 r_s \k0^2/ (1-1.8 \k0^{0.6} + 19 \k0^{1.25})$ accurate (with
$<4$\% error for $\k0 \la 6$).  Similarly for the NFW lenses, we use
$(\beta_{rad} D_l)=-4 r_s \k0 (1-0.4 \k0^{0.1} + 0.5
\k0^{0.2})/\exp[(3+\k0^{-1})/2]$ for
$\sigma_{\rm lens}=\pi (\beta_{rad} D_l)^2$, and $(\theta_{tan} D_l) =
2 r_s (1-0.7 \k0^{0.5} + 1.35 \k0^{0.85})/ \exp[(1+\k0^{-1})/2]$ for
$\theta=2\theta_{tan}$.

We compute the magnification bias $B$ from the fitting formula given
by eq.~(21) in \cite{o02}.  We have tested this formula against
numerical calculations and found good agreement.  A similar fit given
by eq.~(67) in \cite{lo02}, however, substantially underestimates the
bias for $\kappa_0 \la 1$ and overestimates it for $\kappa_0 \ga 1$.
This is because their fit assumed $d\alpha/dx=0$ and therefore
neglected a factor containing $(1-d\alpha/dx)$, where $\alpha$ is the
deflection angle.  We find this not to be a valid assumption in
general.

Fig.~2 compares $dP/d\theta$ for SIS, Moore, and NFW lenses computed
from the halo mass function in eq.~(5), and $dP/d\theta$ for SIS
lenses computed with velocity functions of different slope $\beta$ in
eq.~(3).  It shows that no single mass profile with the halo mass
function can match the $dP/d\theta$ predicted by the observed velocity
function of $\beta\sim -1.3$.  The SIS (solid) and Moore (long-dashed)
profiles predict wrong shapes for $dP/d\theta$, a reflection of the
steeper faint end of the mass function compared with the luminosity
function.  The shape of $dP/d\theta$ for the shallower NFW profile
(short-dashed) resembles more closely the velocity function
prediction, but the lensing amplitude is miniscule.  We note that the
magnification bias $B$ has been included in Fig.~2, which is generally
significantly higher for shallower inner mass profiles, but the
resulting NFW lensing amplitude is still much too low.

\section{Mass vs. Light: Resolution}

To bring the predicted shape for the image separation distribution
from eq.~(\ref{Pmass}) into agreement with eq.~(\ref{Plum}), we
explore the possibility that at a given mass, a fraction of the lenses
is luminous, baryon dominated at the center and has the SIS profile,
while the rest of the lenses is dark matter dominated and has a
shallower inner profile.  Instead of this bimodal model, one can
presumably allow the slope of the inner mass profile to decrease
smoothly with mass.  Observations of lensing galaxies, however,
consistently find that the combined stellar and dark matter mass
profile inside the Einstein radius is well fit by the SIS (e.g., Rix
et al. 1997; Romanowsky \& Kochanek 1999; Cohn et al. 2001; Treu \&
Koopmans 2002).  At the same time, galaxy formation models show that
the energetics of feedback processes are sufficient to expel baryons
in some $<10^{12} M_\odot$ halos.  The bimodal model combining SIS and
dark matter profiles therefore appears physically motivated and will be
used below.

From the solid and long-dashed curves in Fig.~2, we conclude that a
combination of SIS and Moore profiles {\it cannot} reproduce the shape
of $dP/d\theta$ predicted by the velocity function with $\beta \sim
-1.3$.  This is because the inner slope of the Moore profile is close
enough to SIS that the two predict similar shapes for $dP/d\theta$.
Making some lenses dark with $r^{-1.5}$ inner profile will therefore
not reproduce the monotonically rising $dP/d\theta$ at small $\theta$
for the velocity function.  If halos have the shallower $r^{-1}$
profile, however, the dark lenses will have negligible lensing optical
depth compared with the SIS (short-dashed vs. solid in Fig.~2).  We
can then match the two predictions by parameterizing the fraction of
SIS halos at a given mass with
\begin{equation}
   f_{SIS}(M) \propto { (M/M_c)^{\eta_1+\eta_2} \over
   e^{(M/M_c)^{\eta_2}}-1 } \,,
\label{frac}
\end{equation}
which grows as a power law, $f_{SIS} \sim (M/M_c)^{\eta_1}$, at small
$M$ and falls exponentially at large $M$.  This makes the halos mostly
SIS on galactic mass scale $M_c$ where baryon dissipation is
important, and mostly NFW on cluster and sub-galactic scales where
dark matter dominates the potential.  Our main interest here is in
determining the slope $\eta_1$, which has the convenient property that
it depends only on the faint end slope $\beta$ and not on other
parameters in the velocity function in eq.~(\ref{vel}).  It also gives
a simple parameterization of the importance of feedback processes on
the density profile as a function of halo mass.  We note that since
the relation between the image separation $\theta$ and halo mass $M$
is redshift-dependent, the factor $f_{SIS}$ must be included inside
the integral of eq.~(5).  We do not consider explicit redshift
dependence in $f_{SIS}$ here, which can be put in at the expense of
introducing more parameters.  Future work combining $f_{SIS}$
determined from lensing with galaxy formation models and simulations
may offer useful constraints on the time evolution of $f_{SIS}$.

Fig.~3 shows the excellent agreement between the two predictions for
the shape of $dP/d\theta$ for four faint-end slopes of the velocity
function.  The required $\eta_1$ in eq.~(\ref{frac}) is $\approx
0.85$, 0.75, 0.53, and 0.2 for $\beta=-1, -1.3, -2$, and $-3$,
respectively.  The other two parameters $M_c$ and $\eta_2$ depend on
the shape of the velocity function.  For $v_*=250$ km s$^{-1}$ and
$n=2.5$ from the SSRS2 sample \citep{G00}, we find a good match with
$M_c\approx 4.5\times 10^{11} M_\odot$ and $\eta_2\approx 0.72$.  We
also find it necessary to lower the overall amplitude of $dP/d\theta$
for the dashed curves in Fig.~3 by $\sim 20$\% to 40\% (for normalized
$f_{SIS}$ shown in Fig.~4) to match the dotted curves.  We have not
attempted to fine tune it since the amplitude of $dP/d\theta$ depends
on several uncertain parameters, e.g., the source redshift
distribution, the normalization and redshift evolution of the
luminosity function, and the precise value of $\gamma_v=v_c/v_{vir}$
for SIS halos.  Instead we have focused on the constraints from the
shape of $dP/d\theta$.

Fig.~4 shows the required $f_{SIS}(M)$ for each of the four velocity
functions in Fig.~3.  Our $f_{SIS}$ at the high mass end for
$\beta=-1.3$ agrees well with the result from \cite{oguri02}, which
used $f_{SIS}(M)=\exp[1-(M/M_h)^{\delta_h}]$ for $M>M_h$ following
\cite{k01}, and set $f_{SIS}=1$ for $M<M_h$, i.e., it ignored the dark
lens fraction for small masses.  By contrast, our form of $f_{SIS}$ in
eq.~(\ref{frac}) is a smooth function and takes into account all
masses.

Fig.~4 illustrates that a very steep faint-end slope ($\beta\sim -4$)
for the velocity function will be required if low-mass halos all have
the SIS profile.  Current galaxy surveys favor a much shallower faint
end slope of $\beta\sim -1.3$.  We thus conclude that the percentage
of halos that can have the SIS profile must decrease rapidly with
decreasing halo mass below $10^{12} M_\odot$.  This implies that
feedback processes are increasingly effective in reducing the baryon
content in small objects, a trend consistent with semi-analytic galaxy
formation models.  Moreover, the halos that have non-SIS profiles must
have an inner density of $\rho\sim r^{-1}$ or shallower so as to
contribute negligible lensing optical depth.  (The possibly shallow
profiles of dwarf galaxies will therefore not affect the lensing
predictions here.)  The steeper $\rho\sim r^{-1.5}$ would predict a
shape for the lensing image separation different from the observed
luminosity function.

I have enjoyed discussions with Paul Schechter, Joe Silk, Aveshi
Dekel, David Rusin, Dragan Huterer, and David Hogg.  This work is
supported in part by an Alfred P. Sloan Foundation Fellowship, a
Cottrell Scholars Award from the Research Corporation, and NASA grant
NAG 5-12173.

\clearpage
\begin{figure*}
\begin{center}
\begin{tabular}{c}
\epsscale{.8}
\plotone{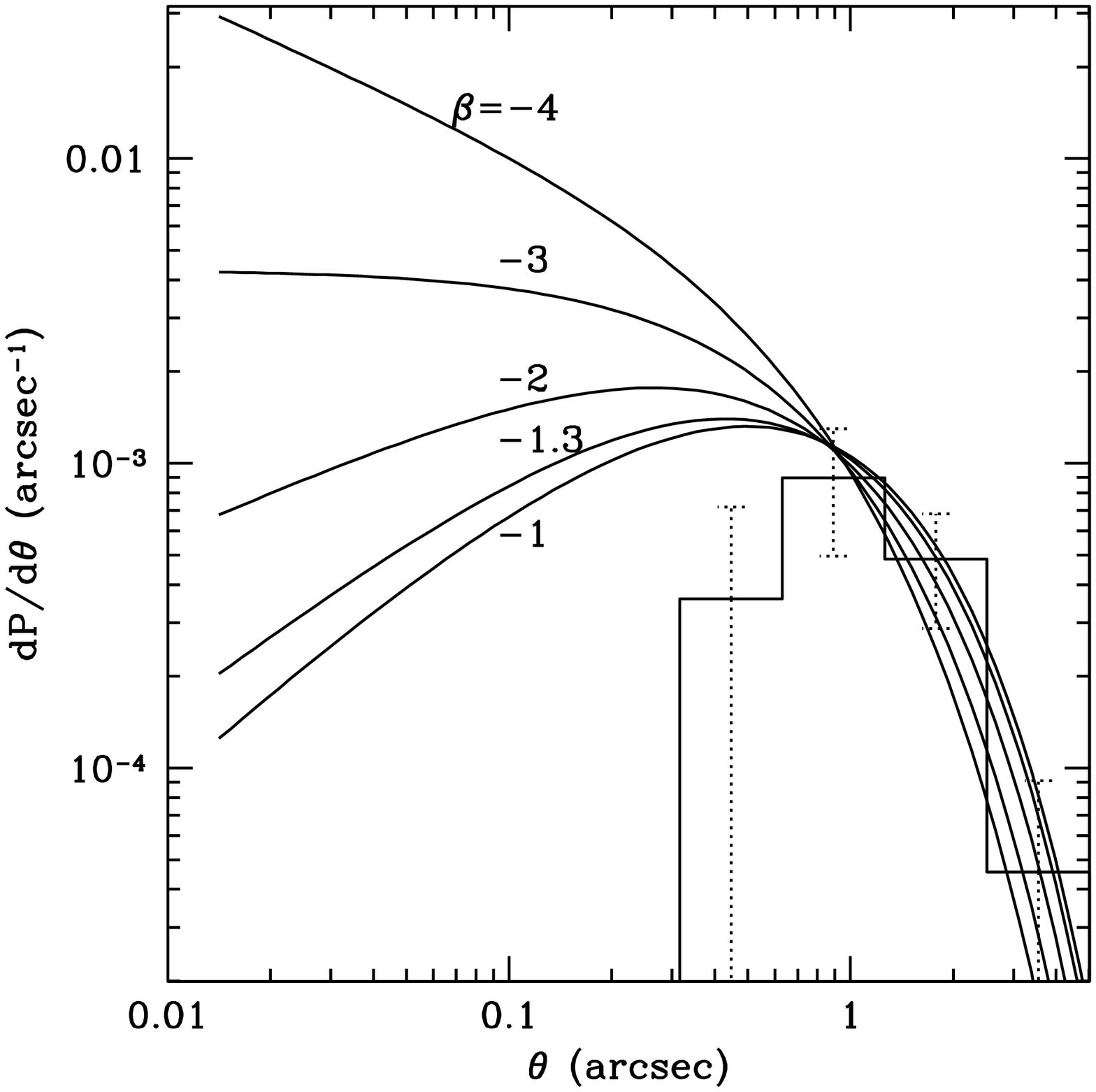}
\end{tabular}
\figurenum{1}\label{f1}
\caption{ Image separation distribution $dP/d\theta$ calculated from
eq.~(3) using the galaxy velocity function in eq.~(2) and SIS profile.
For small $\theta$, the slope of $dP/d\theta$ only depends on the
faint-end slope $\beta$ of the velocity function, changing from
positive to negative when the velocity function steepens beyond
$\beta=-3$.  Other velocity function parameters used here are
$v_*=250$ km s$^{-1}$, $n=2.5$, $\psi_*=0.073 h^3$ Mpc$^{-3}$ from the
SSRS2 sample in Gonzalez et al. (2000).  The sources are put at the
mean redshift 1.27 of the CLASS radio sources with a power-law flux
distribution of slope -2.1.  The histogram shows the 13 CLASS lenses
and the $1\sigma$ poisson errors (Browne et al. 2002).}
\end{center}
\end{figure*}

\clearpage

\begin{figure*}
\begin{center}
\begin{tabular}{c}
\epsscale{.8}
\plotone{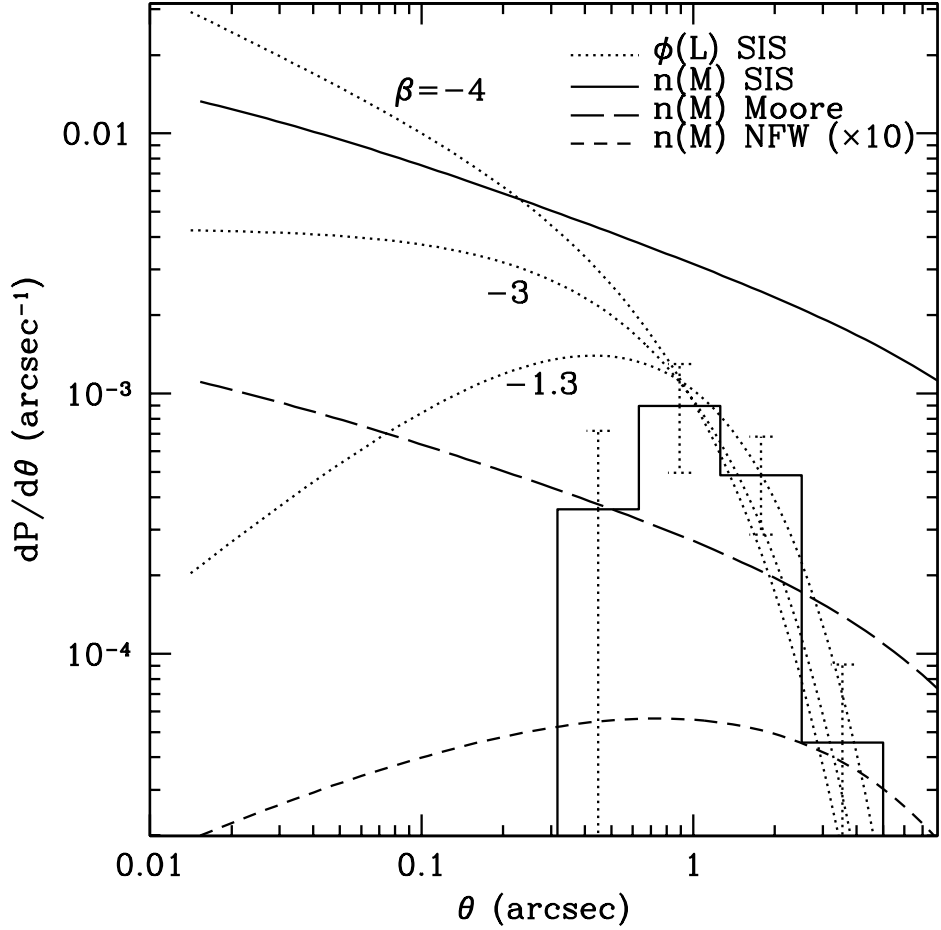}
\end{tabular}
\figurenum{2}\label{f2}
\caption{ Image separation distribution $dP/d\theta$ calculated from
galaxy velocity function (eq.~(3); dotted) vs. mass function
(eq.~(5)).  The result depends strongly on the choice of the function
and the assumed lens mass profile.  The lensing amplitude decreases
rapidly as the inner mass profile is lowered from $-2$ (SIS; solid),
to $-1.5$ (Moore; long-dashed), to $-1$ (NFW raised by a factor of 10
to fit in the plot; short-dashed).  Note that no single halo profile
can bring the mass function prediction for $dP/d\theta$ into agreement
with the velocity function prediction.  The histogram shows CLASS data
as in Fig.~1.}
\end{center}
\end{figure*}

\clearpage

\begin{figure*}
\begin{center}
\begin{tabular}{c}
\epsscale{.8}
\plotone{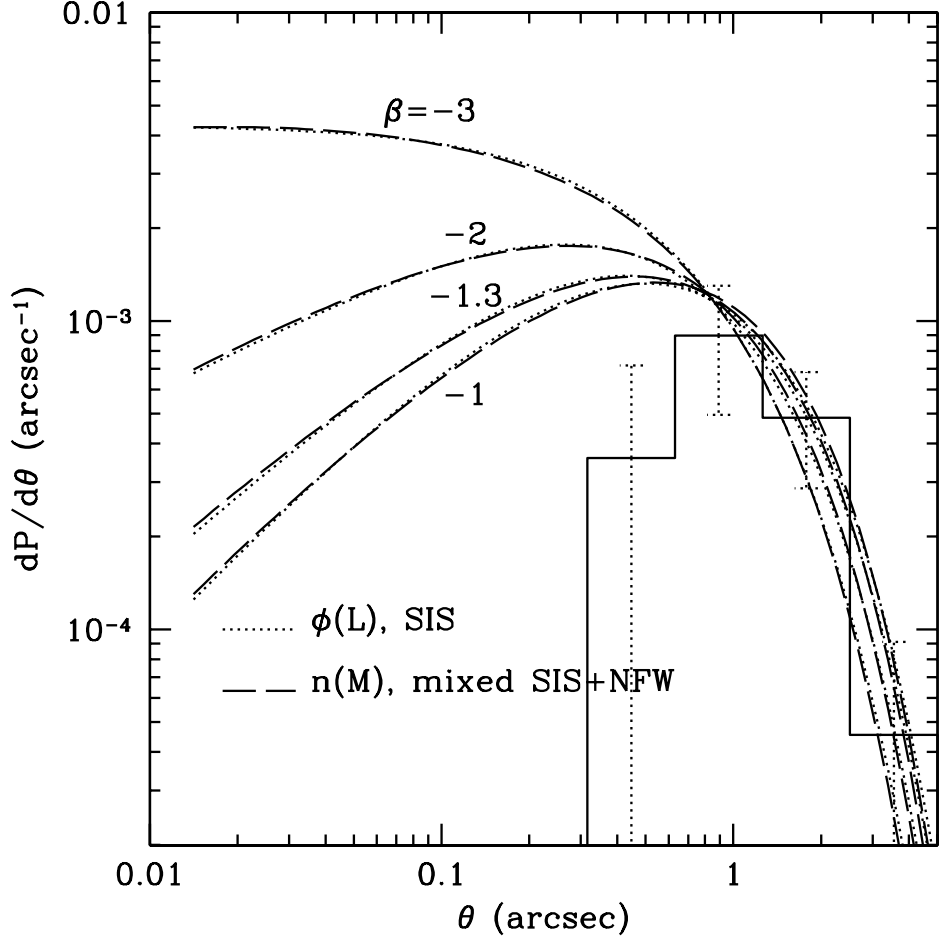}
\end{tabular}
\figurenum{3}\label{f3}
\caption{ Different predictions from velocity vs. mass function in
Fig.~2 can be brought into agreement if at a given mass, a fraction
$f_{SIS}(M)$ (eq.~(6)) of the dark matter halos is assigned SIS and
the rest NFW.  The predicted $dP/d\theta$ from the velocity function
(dotted) and the mass function (dashed) then agree very well for
suitable choices of parameters $\eta_1, \eta_2$ and $M_c$ for
$f_{SIS}$ (see text).  The histogram shows CLASS data as in Fig.~1.}
\end{center}
\end{figure*}

\clearpage

\begin{figure*}
\begin{center}
\begin{tabular}{c}
\epsscale{.8}
\plotone{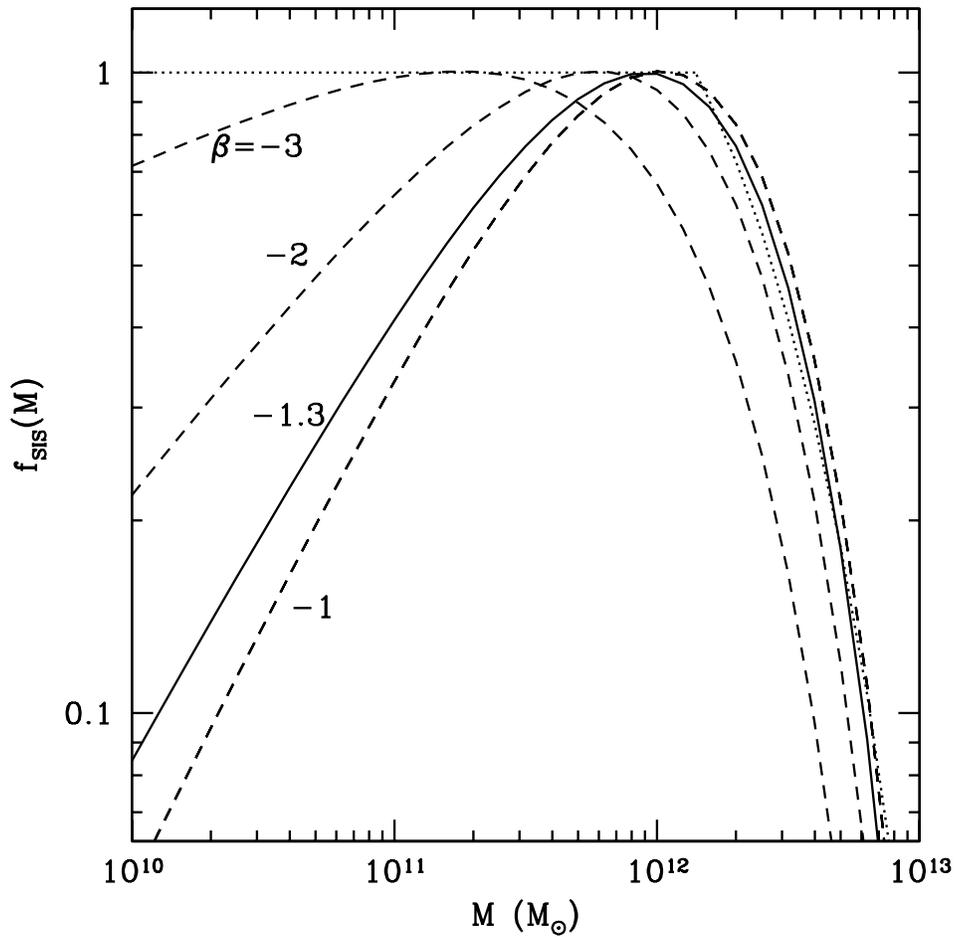}
\end{tabular}
\figurenum{4}\label{f4}
\caption{ The fraction $f_{SIS}(M)$ of halos with SIS profiles needed
for the consistent predictions in Fig.~3.  As the faint-end slope
$\beta$ of the velocity function steepens, a larger fraction of
low-mass halos is allowed to be SIS.  Galaxy surveys favor $\beta\sim
-1.3$ (solid), requiring $f_{SIS}\sim M^{\eta_1}$ with $\eta_1\approx
0.75$ below $10^{12} M_\odot$.
The dotted curve shows Oguri's result (2002), which agrees well
with our solid curve at large mass but assumes $f_{SIS}=1$ at
small mass.}
\end{center}
\end{figure*}

\end{document}